\documentclass{article}
\usepackage{spconf,amsmath,epsfig}
\usepackage[english]{babel} 
\usepackage{amsmath}                
\usepackage{amsfonts}               
\usepackage{amssymb}                
\usepackage{amsopn}                 
\allowdisplaybreaks[3] 
\usepackage{amsthm}                
\usepackage{bbm}                    
\usepackage{mathrsfs}               
\usepackage[retainorgcmds]{IEEEtrantools} 
\usepackage{calc}                   
\usepackage{epsfig}                
\usepackage{psfrag}                 
\usepackage[dvips]{color}           
\usepackage{fancyhdr}               
\usepackage{verbatim}               
\usepackage{exscale}                
\usepackage{macros}
\newtheorem{proposition}{Proposition}
\newtheorem{remark}{Remark}

\newcommand{\Ave}{\mathcal{E}}
\newcommand{\Peak}{\mathcal{A}}
\renewcommand{\Poisson}{\mathcal{P}}
\newcommand{\Qb}{Q^{\textnormal{b}}}

\title{THE POISSON CHANNEL AT LOW INPUT POWERS}
\threeauthors
  {Amos~Lapidoth, Ligong~Wang}
   {
   ETH Zurich\\
   Zurich, Switzerland\\
   \texttt{\{lapidoth,wang\}@isi.ee.ethz.ch} }
  {Jeffrey~H.~Shapiro}
   {MIT\\
   Cambridge, MA, USA\\
   \texttt{jhs@mit.edu}}
   {Vinodh~Venkatesan}
   {IBM Research\\
   Zurich, Switzerland\\
   \texttt{ven@zurich.ibm.com}}
   
\begin{document}
	
\maketitle

\begin{abstract}
	The asymptotic capacity at low input powers of an average-power limited or an average- and peak-power limited discrete-time Poisson channel is considered. For a Poisson channel whose dark current is zero or decays to zero linearly with its average input power~$\Ave$, capacity scales like~$\Ave\log\frac{1}{\Ave}$ for small~$\Ave$. For a Poisson channel whose dark current is a nonzero constant, capacity scales, to within a constant, like~$\Ave\log\log\frac{1}{\Ave}$ for small~$\Ave$.
\end{abstract}


\section{introduction}

We consider the discrete-time memoryless Poisson channel whose input $x$ is in the set $\Reals_0^+$ of nonnegative reals and whose output $y$ is in the set $\Integers_0^+$ of nonnegative integers. Conditional on the input $X=x$, the output $Y$ has a Poisson distribution of mean $(\lambda+x)$ where $\lambda\ge 0$ is called the \emph{dark current}. We denote the Poisson distribution of mean $\xi$ by $\Poisson_\xi(\cdot)$ so
\begin{equation*}
	\Poisson_\xi(y) = e^{-\xi} \frac{\xi^y}{y!},\quad y\in\Integers_0^+.
\end{equation*}
With this notation the channel law $W(\cdot|\cdot)$ is given by
\begin{equation}\label{eq:channel_law}
	W(y|x) = \Poisson_{\lambda+x}(y),\quad x\in\Reals_0^+,y\in\Integers_0^+.
\end{equation}

This channel is often used to model pulse-amplitude modulated optical communication 
with a direct-detection receiver \cite{shamai90}. Here the input~$x$ is proportional to the product of  the transmitted light intensity by the pulse duration; the output $y$ models the number of photons arriving at the receiver during the pulse duration; and $\lambda$ models the average number of extraneous counts that appear in $y$ in addition to those associated with the illumination $x$.

The \emph{average-power constraint}\footnote{The word ``power'' here has the meaning ``average number of photons transmitted per channel use.'' If we denote by $P$ the standard ``power'' in physics, namely, energy per unit time (in watts), then the notation of ``power'' in this paper is really $\eta PT/\hbar \omega$, where $\eta$ is the detector's quantum efficiency, $T$ is the pulse duration (in sec), and $\hbar\omega$ is the photon energy (in joules) at the operating frequency $\omega$ (in rad/sec).} on the input is
\begin{equation}\label{eq:ave}
	\E{X} \le \Ave ,
\end{equation}
where $\Ave >0$ is the maximum allowed average power. 

The \emph{peak-power constraint} on the input is that with probability one
\begin{equation}\label{eq:peak}
	X \le \Peak.
\end{equation}
When no peak-power constraint is imposed, we write $\Peak = \infty$.

No analytic expression for the capacity of the Poisson channel is known. In~\cite{shamai90} Shamai showed that capacity-achieving input distributions are discrete whose numbers of mass points depend on $\Ave$ and $\Peak$. In~\cite{lapidothmoser03_4,lapidothmoser07_sub} Lapidoth and Moser derived the asymptotic capacity of the Poisson channel in the regime where both the average and peak powers tend to infinity with their ratio fixed.

In the present paper, we seek the asymptotic capacity of the Poisson channel when the average input power tends to zero. The peak-power constraint, when considered, is held constant and hence does not tend to zero with the average power. We consider two different cases for the dark current~$\lambda$. The first case is when the dark current tends to zero proportionally with the average power. This corresponds to the wide-band regime where the pulse duration tends to zero. The second case is when the dark current is constant. This corresponds to the regime where the transmitter is weak.

Our lower bounds on channel capacity are all based on binary inputs. In some cases we show that this is asymptotically optimal. Our upper bounds are derived using the duality expression (see \cite{lapidothmoser03_3} and references therein). An efficient way to compute asymptotic capacities at low average input powers is to compute the \emph{capacity per unit cost} \cite{verdu90}. However, we shall see that, apart from one case (Equation \eqref{eq:dark_peak}), the capacity per unit cost does not exist, namely, the capacity tends to zero more slowly than linearly with the average power.

Among the results in this paper, the special case of zero dark current has been derived independently in \cite{venkatesan08,martinez08}.

The rest of the paper is arranged as follows: in Section~\ref{sec:main} we state the results of this paper; in Section~\ref{sec:lower} we prove the lower bounds; and in Section~\ref{sec:upper} we sketch the proofs for the upper bounds.


\section{results}\label{sec:main}

Let $C(\lambda, \Ave, \Peak)$ denote the capacity of the Poisson channel with dark current $\lambda$ under Constraints~\eqref{eq:ave} and~\eqref{eq:peak}
\begin{equation*}
	C(\lambda, \Ave, \Peak) = \sup I(X;Y)
\end{equation*}
where the supremum is over all input distributions satisfying~\eqref{eq:ave} and~\eqref{eq:peak}.

When $\lambda$ is proportional to~$\Ave$, the asymptotic capacity of the Poisson channel as~$\Ave\downarrow 0$ is given in the following proposition. Note that this also includes the case where the dark current is the constant zero.  





\begin{proposition}[Dark Current Proportional to $\Ave$]
\label{pro:prop}
	For any $c\ge0$ and $\Peak \in (0,\infty]$,
	\begin{equation*}
		\lim_{\Ave\downarrow 0} \frac{C(c\Ave, \Ave, \Peak)}{\Ave\log\frac{1}{\Ave}} = 1.
	\end{equation*}
\end{proposition}

Recall that, for any $\alpha, \beta>0$, the sum of two independent random variables with the Poisson distributions $\Poisson_\alpha(\cdot)$ and $\Poisson_\beta(\cdot)$ has the Poisson distribution~$\Poisson_{\alpha+\beta}(\cdot)$. Thus, we can produce any Poisson channel with nonzero dark current by adding noise to a Poisson channel with zero dark current. Consequently,
\begin{equation*}
	C(0,\Ave,\Peak) \ge C(c\Ave,\Ave,\Peak),\quad c, \Ave, \Peak>0.
\end{equation*}
Thus, to prove Proposition~\ref{pro:prop}, we only need to show the following two bounds:
\begin{IEEEeqnarray}{rCl}
	\varliminf_{\Ave\downarrow 0} \frac{C(c\Ave, \Ave, \Peak)}{\Ave\log\frac{1}{\Ave}} & \ge &1,\quad c>0,\Peak\in(0,\infty], \label{eq:prop_lower}\\
	\varlimsup_{\Ave\downarrow 0} \frac{C(0, \Ave, \Peak)}{\Ave\log\frac{1}{\Ave}} & \le &1,\quad\Peak\in(0,\infty].\label{eq:zero_upper}	
\end{IEEEeqnarray}
We shall prove~\eqref{eq:prop_lower} in Section~\ref{sub:prop_lower} and shall sketch a proof for~\eqref{eq:zero_upper} in Section~\ref{sub:zero_upper}.

\begin{remark}
	The bound \eqref{eq:zero_upper}  can also be derived by noting that the capacity of the Poisson channel with zero dark current under an average-power constraint only is upper-bounded by the capacity of the pure-loss bosonic channel, and by using the explicit formula \cite{giovanettiguha04}
	\begin{equation}\label{eq:bosonic}
		C_{\textnormal{bosonic}}(\Ave) = (1+\Ave)\log(1+\Ave) - \Ave\log\Ave
	\end{equation}
	of the latter.  
\end{remark}
\begin{remark}
	Because the pure-loss bosonic channel with coherent input states and direct detection reduces to a Poisson channel, the lower bound~\eqref{eq:prop_lower} and the achievability of its left-hand side using binary signaling combine with~\eqref{eq:bosonic} to show that the asymptotic (quantum-receiver) capacity of the pure-loss bosonic channel is achievable with binary modulation (on-off keying) and direct detection.
\end{remark}

For a Poisson channel with constant nonzero dark current, we have the following result.
\begin{proposition}[Constant Nonzero Dark Current]
\label{pro:dark}
	For any $\lambda>0$,
	\begin{equation}\label{eq:dark_peak}
		\lim_{\Ave\downarrow 0} \frac{C(\lambda, \Ave, \Peak)}{\Ave} = \left(1+\frac{\lambda}{\Peak}\right)\log \left(1+\frac{\Peak}{\lambda}\right) - 1, \quad \Peak < \infty,
	\end{equation}
	and
	\begin{equation}\label{eq:dark}
		\frac{1}{2} \le \varliminf_{\Ave \downarrow 0} \frac{C(\lambda, \Ave, \infty)}{\Ave \log\log\frac{1}{\Ave }} \le \varlimsup_{\Ave \downarrow 0} \frac{C(\lambda, \Ave, \infty)}{\Ave \log\log\frac{1}{\Ave }}\le 2.
	\end{equation}
\end{proposition}

The proof of~\eqref{eq:dark_peak} is a simple application of the formula for capacity per unit cost \cite[Theorem 2]{verdu90}. The proof of the lower bound in~\eqref{eq:dark} is in Section~\ref{sub:dark_lower}; and a sketch of the proof of the upper bound in~\eqref{eq:dark} is in Section~\ref{sub:dark_upper}.


\section{the lower bounds}\label{sec:lower}

The achievability results in this section are obtained by choosing binary input distributions and then computing the mutual informations. We denote by $\Qb$ the binary distribution
\begin{equation}\label{eq:input}
	X = \begin{cases} 0,& \textnormal{w.p. }(1-p),\\ \zeta, & \textnormal{w.p. }p,\end{cases}
\end{equation}
where $\zeta>0$, $p\in(0,1)$.
If we choose the parameters $\zeta$ and $p$ in such a way that Constraints~\eqref{eq:ave} and~\eqref{eq:peak} are satisfied, then
\begin{equation}\label{eq:lower_general}
	C(\lambda, \Ave, \Peak) \ge I(\Qb, W).
\end{equation}

\subsection{Dark Current Proportional to $\Ave$}\label{sub:prop_lower}

In this subsection we shall derive Inequality~\eqref{eq:prop_lower}. To this end, we compute the mutual information $I(\Qb,W)$ for input distribution $\Qb$ given by~\eqref{eq:input}:
\begin{IEEEeqnarray}{rCl}
	I(\Qb,W) & = & H(Y) - H(Y|X) \nonumber\\
	& = & -\sum_{y=0}^{\infty} \bigl( (1-p)\Poisson_\lambda(y) + p \Poisson_{\lambda+\zeta}(y) \bigr)\nonumber\\
	&&~~~~~~\cdot \log \bigl( (1-p)\Poisson_\lambda(y) + p \Poisson_{\lambda+\zeta}(y) \bigr)\nonumber\\
	&& {}+(1-p) \sum_{y=0}^\infty \Poisson_\lambda(y) \log \Poisson_\lambda(y) \nonumber\\
	&& {} + p \sum_{y=0}^\infty \Poisson_{\lambda+\zeta}(y) \log \Poisson_{\lambda+\zeta}(y)\nonumber\\
	&=& I_0(\lambda,\zeta,p) + I_1(\lambda,\zeta,p),
	\label{eq:prop_l1}
\end{IEEEeqnarray}
where in the last equality we defined
\begin{IEEEeqnarray*}{rCl}
	I_0(\lambda,\zeta,p) & \triangleq & -\bigl( (1-p) e^{-\lambda} + p e^{-(\lambda+\zeta)} \bigr) \nonumber\\
	&& ~~~~~~\cdot \log \bigl( (1-p) e^{-\lambda} + p e^{-(\lambda+\zeta)} \bigr) \nonumber\\
	&& {} - (1-p)\lambda e^{-\lambda} -p(\lambda+\zeta)e^{-(\lambda+\zeta)},
	\\
	I_1(\lambda,\zeta,p) & \triangleq &  -\sum_{y=1}^{\infty} \bigl( (1-p)\Poisson_\lambda(y) + p \Poisson_{\lambda+\zeta}(y) \bigr) \nonumber\\
	&& ~~~~~~\cdot\log \bigl( (1-p)\Poisson_\lambda(y) + p \Poisson_{\lambda+\zeta}(y) \bigr)\nonumber\\
	&& {} +(1-p) \sum_{y=1}^\infty \Poisson_\lambda(y) \log \Poisson_\lambda(y) \nonumber\\
	&& {} + p \sum_{y=1}^\infty \Poisson_{\lambda+\zeta}(y) \log \Poisson_{\lambda+\zeta}(y).
\end{IEEEeqnarray*}
Note that in the above decomposition we took out the terms corresponding to $y=0$ in all three summations to form $I_0(\lambda,\zeta,p)$ and collected the remaining terms in $I_1(\lambda,\zeta,p)$.

We lower-bound $I_0(\lambda,\zeta,p)$ as
\begin{IEEEeqnarray}{rCl}
	I_0(\lambda,\zeta,p) & \ge & 0 - (1-p) \lambda e^{-\lambda} - p(\lambda+\zeta) e^{-(\lambda+\zeta)} \nonumber\\
	& \ge & -\lambda - p(\lambda+\zeta). \label{eq:prop_l2}
\end{IEEEeqnarray}

We lower-bound $I_1(\lambda,\zeta,p)$ as
\begin{IEEEeqnarray}{rCl}
	\lefteqn{I_1(\lambda,\zeta,p)}\nonumber\\
	& = & - \sum_{y=1}^\infty \bigl( (1-p)\Poisson_\lambda(y) + p \Poisson_{\lambda+\zeta}(y) \bigr) \nonumber\\
	&& ~~~~~~\cdot \log \left((1-p)\frac{\Poisson_\lambda(y)}{\Poisson_{\lambda+\zeta}(y)} + p \right) \nonumber\\
	&& {} +  (1-p) \sum_{y=1}^\infty \Poisson_\lambda(y) \log \frac{\Poisson_\lambda(y)}{\Poisson_{\lambda+\zeta}(y)} \nonumber\\
	& = & - \sum_{y=1}^\infty \bigl( (1-p)\Poisson_\lambda(y) + p \Poisson_{\lambda+\zeta}(y) \bigr) \nonumber\\
	&&~~~~~~\cdot\left(\log p + \log \left(1+\frac{1-p}{p}\frac{\Poisson_\lambda(y)}{\Poisson_{\lambda+\zeta}(y)} \right)\right) \nonumber\\
	&&{} + (1-p) \sum_{y=1}^\infty \Poisson_\lambda(y) \log \frac{e^{-\lambda} \frac{\lambda^y}{y!}}{e^{-(\lambda+\zeta)}\frac{(\lambda+\zeta)^y}{y!}}\nonumber\\
	& = & - \sum_{y=1}^\infty \bigl( (1-p)\Poisson_\lambda(y) + p \Poisson_{\lambda+\zeta}(y) \bigr) \nonumber\\
	&& ~~~~~~\cdot \Bigg(\log p + \underbrace{\log \left(1+\frac{1-p}{p}\frac{\Poisson_\lambda(y)}{\Poisson_{\lambda+\zeta}(y)} \right)}_{\le \frac{1-p}{p}\frac{\Poisson_\lambda(y)}{\Poisson_{\lambda+\zeta}(y)}}\Bigg) \nonumber\\
	&& {}+ (1-p)\zeta\underbrace{\sum_{y=1}^\infty \Poisson_\lambda(y)}_{=1-e^{-\lambda}}  + (1-p) \log\frac{\lambda}{\lambda+\zeta}\underbrace{\sum_{y=1}^\infty \Poisson_\lambda(y) y }_{=\lambda}\nonumber\\
	& \ge & - \underbrace{\sum_{y=1}^\infty \bigl( (1-p)\Poisson_\lambda(y) + p \Poisson_{\lambda+\zeta}(y) \bigr)}_{=(1-p)(1-e^{-\lambda}) + p(1-e^{-(\lambda+\zeta)})} \log p \nonumber\\
	&& {} - \sum_{y=1}^\infty \bigl( (1-p)\Poisson_\lambda(y) + p \Poisson_{\lambda+\zeta}(y) \bigr)\frac{1-p}{p}\frac{\Poisson_\lambda(y)}{\Poisson_{\lambda+\zeta}(y)} \nonumber\\
	&& {}+ (1-p)(1-e^{-\lambda})\zeta - (1-p)\lambda\log\left(1+\frac{\zeta}{\lambda}\right) \nonumber\\
	& = & \bigl( (1-p)(1-e^{-\lambda}) + p(1-e^{-(\lambda+\zeta)})\bigr) \log \frac{1}{p} \nonumber\\
	&& {}-\frac{(1-p)^2}{p} \sum_{y=1}^\infty \underbrace{\frac{\left(\Poisson_\lambda(y)\right)^2}{\Poisson_{\lambda+\zeta}(y)}}_{=e^{\frac{\zeta^2}{\lambda+\zeta}} \Poisson_{\frac{\lambda^2}{\lambda+\zeta}}(y)} - (1-p) \underbrace{\sum_{y=1}^\infty \Poisson_\lambda(y)}_{=1-e^{-\lambda}} \nonumber\\
	&& {}+ (1-p)(1-e^{-\lambda}) \zeta - (1-p) \lambda \log\left(1+\frac{\zeta}{\lambda}\right)\nonumber\\
	& = & \bigl( (1-p)(1-e^{-\lambda}) + p(1-\underbrace{e^{-(\lambda+\zeta)}}_{\le e^{-\zeta}})\bigr) \log \frac{1}{p} \nonumber\\
	&& {}-\underbrace{\frac{(1-p)^2}{p}}_{\le \frac{1}{p}} \underbrace{e^{\frac{\zeta^2}{\lambda+\zeta}}}_{\le e^\zeta} \underbrace{\left(1- e^{-\frac{\lambda^2}{\lambda+\zeta}}\right)}_{\le \frac{\lambda^2}{\lambda+\zeta} \le \frac{\lambda^2}{\zeta}} - \underbrace{(1-p)}_{\le 1}\underbrace{(1-e^{-\lambda})}_{\le \lambda} \nonumber\\
	&& {}-\underbrace{(1-p)}_{\le 1}\underbrace{(1-e^{-\lambda})}_{\le \lambda}\zeta - (1-p) \lambda \log \left(1+\frac{\zeta}{\lambda}\right) \nonumber\\
	& \ge & (1-p)(1-e^{-\lambda})\log\frac{1}{p} + p(1-e^{-\zeta})\log\frac{1}{p}\nonumber\\
	&& {} - \frac{1}{p}\frac{\lambda^2}{\zeta} e^\zeta -\lambda -\lambda \zeta - (1-p) \lambda \log \left(1+\frac{\zeta}{\lambda}\right). \label{eq:prop_l3}
\end{IEEEeqnarray}

Choose any $\zeta\in (0, \Peak]$ and, for small enough $\Ave$, let $p=\Ave/\zeta$. Then the distribution~\eqref{eq:input} satisfies both Constraints~\eqref{eq:ave} and~\eqref{eq:peak}. Let $\lambda=c\Ave$. Using~\eqref{eq:prop_l2} we can bound the asymptotic behavior of $I_0(\lambda,\zeta,p)$ as
\begin{equation}
	\varliminf_{\Ave\downarrow 0} \frac{I_0\left(c\Ave, \zeta, \frac{\Ave}{\zeta}\right)}{\Ave\log\frac{1}{\Ave}} \ge -\lim_{\Ave\downarrow 0}\frac{c\Ave}{\Ave\log\frac{1}{\Ave}} - \lim_{\Ave \downarrow 0} \frac{\frac{\Ave}{\zeta} \left(c \Ave+ \zeta\right)}{\Ave\log\frac{1}{\Ave}} = 0.\label{eq:prop_l4}
\end{equation}
Similarly, using~~\eqref{eq:prop_l3} we can bound the asymptotic behavior of $I_1(\lambda,\zeta,p)$ as
\begin{equation}
	\varliminf_{\Ave\downarrow 0} \frac{I_1\left(c\Ave, \zeta, \frac{\Ave}{\zeta}\right)}{\Ave\log\frac{1}{\Ave}}
	\ge\frac{1-e^{-\zeta}}{\zeta}.
	\label{eq:prop_l5}
\end{equation}
Combining~\eqref{eq:lower_general},~\eqref{eq:prop_l1},~\eqref{eq:prop_l4}, and~\eqref{eq:prop_l5} we obtain
\begin{equation}\label{eq:prop_l6}
	\varliminf_{\Ave\downarrow 0} \frac{C(c\Ave, \Ave, \Peak)}{\Ave\log\frac{1}{\Ave}} \ge \frac{1-e^{-\zeta}}{\zeta},\quad \textnormal{for all }\zeta\in(0, \Peak].
\end{equation}
We can make the right-hand side (RHS) of~\eqref{eq:prop_l6} arbitrarily close to $1$ by choosing arbitrarily small positive values for $\zeta$. Thus we obtain~\eqref{eq:prop_lower}.


\subsection{Constant Nonzero Dark Current}\label{sub:dark_lower}
In this subsection we shall prove the first inequality in~\eqref{eq:dark}. To this end, we lower-bound on the mutual information~$I(\Qb,W)$ for the input distribution~\eqref{eq:input} as follows:
\begin{IEEEeqnarray}{rCl}
 	\lefteqn{I(\Qb,W)}\nonumber\\
	&=& H(Y) - H(Y|X) \nonumber\\
	&=& - \sum_{y=0}^{\infty} \left( (1-p) \Poisson_{\lambda}(y) + p \Poisson_{\lambda+\zeta}(y) \right) \nonumber\\
	&& ~~~~~~\cdot \log \left( (1-p) \Poisson_{\lambda}(y) + p \Poisson_{\lambda+\zeta}(y) \right) \nonumber \\
	& & {} + (1-p) \sum_{y=0}^{\infty} \Poisson_{\lambda}(y) \log \Poisson_{\lambda}(y) \nonumber\\
	&& {} + p \sum_{y=0}^{\infty} \Poisson_{\lambda+\zeta}(y) \log \Poisson_{\lambda+\zeta}(y) \nonumber\\
	&=& {} - p \sum_{y=0}^{\infty} \Poisson_{\lambda+\zeta}(y) \log \left( (1-p) \frac{\Poisson_{\lambda}(y)}{\Poisson_{\lambda+\zeta}(y)} + p \right) \nonumber \\
	& & {} - (1-p) \sum_{y=0}^{\infty} \Poisson_{\lambda}(y) \log \left( (1-p) + p \frac{\Poisson_{\lambda+\zeta}(y)}{\Poisson_{\lambda}(y)} \right) \nonumber\\
	&=& {} - p \sum_{y=0}^{\infty} \Poisson_{\lambda+\zeta}(y) \nonumber\\
	&& ~~~~~~\cdot \Biggl( \log \frac{\Poisson_{\lambda}(y)}{\Poisson_{\lambda+\zeta}(y)}  + \log \left( (1-p)  + p \frac{\Poisson_{\lambda+\zeta}(y)}{\Poisson_{\lambda}(y)} \right) \Biggr) \nonumber \\
	& & {} - (1-p) \sum_{y=0}^{\infty} \Poisson_{\lambda}(y) \log \left( (1-p) + p \frac{\Poisson_{\lambda+\zeta}(y)}{\Poisson_{\lambda}(y)} \right) \nonumber\\
	&=& p \sum_{y=0}^{\infty} \Poisson_{\lambda+\zeta}(y) \log \frac{\Poisson_{\lambda+\zeta}(y)}{\Poisson_{\lambda}(y)} - \sum_{y=0}^{\infty} \Bigg( \underbrace{(1-p) \Poisson_{\lambda}(y)}_{\ge 0} \nonumber \\
	& & {}  + \underbrace{p \Poisson_{\lambda+\zeta}(y)}_{\ge 0} \Bigg) \underbrace{\log \left( (1-p) + p \frac{\Poisson_{\lambda+\zeta}(y)}{\Poisson_{\lambda}(y)} \right)}_{\le \log \left( 1 + p\frac{\Poisson_{\lambda+\zeta}(y)}{\Poisson_{\lambda}(y)} \right) \le p \frac{\Poisson_{\lambda+\zeta}(y)}{\Poisson_{\lambda}(y)}} \nonumber\\
	&\ge& p \sum_{y=0}^{\infty} \Poisson_{\lambda+\zeta}(y) \log \frac{\Poisson_{\lambda+\zeta}(y)}{\Poisson_{\lambda}(y)} \nonumber \\
	& & {} - \sum_{y=0}^{\infty} \Big( (1-p) \Poisson_{\lambda}(y) + p \Poisson_{\lambda+\zeta}(y) \Big) p \frac{\Poisson_{\lambda+\zeta}(y)}{\Poisson_{\lambda}(y)} \nonumber\\
	&=& p \sum_{y=0}^{\infty} \Poisson_{\lambda+\zeta}(y) \log \left( \frac{e^{-(\zeta + \lambda)} \frac{(\zeta + \lambda)^y}{y!}}{e^{-\lambda} \frac{\lambda^y}{y!}} \right) \nonumber\\
	&& {} - (1-p)p \sum_{y=0}^{\infty} \Poisson_{\lambda}(y) \frac{\Poisson_{\lambda+\zeta}(y)}{\Poisson_{\lambda}(y)} \nonumber \\
	& & {} - p^2 \sum_{y=0}^{\infty} \Poisson_{\lambda+\zeta}(y) \frac{\Poisson_{\lambda+\zeta}(y)}{\Poisson_{\lambda}(y)} \nonumber\\
	&=& p \sum_{y=0}^{\infty} \Poisson_{\lambda+\zeta}(y) \log \left( e^{-\zeta} \left( 1 + \frac{\zeta}{\lambda} \right)^y \right) \nonumber\\
	&& {} - (1-p)p \underbrace{\sum_{y=0}^{\infty} \Poisson_{\lambda+\zeta}(y)}_{=1} - p^2 \sum_{y=0}^{\infty} \frac{\left( e^{-(\zeta + \lambda)} \frac{(\zeta + \lambda)^y}{y!} \right)^2}{e^{-\lambda} \frac{\lambda^y}{y!}} \nonumber\\
	&=& p \sum_{y=0}^{\infty} \Poisson_{\lambda+\zeta}(y) \left( -\zeta + y \log \left( 1 + \frac{\zeta}{\lambda} \right) \right) - (1-p)p \nonumber \\
	& & {} - p^2 \underbrace{\left(\sum_{y=0}^{\infty} e^{-(\lambda+2\zeta)} \frac{\left( \lambda + 2\zeta + \frac{\zeta^2}{\lambda} \right)^y}{y!} e^{-\frac{\zeta^2}{\lambda}}\right)}_{=\sum_{y=0}^{\infty} \Poisson_{\frac{\lambda + 2\zeta + \zeta^2}{\lambda} }(y) = 1} e^{\frac{\zeta^2}{\lambda}} \nonumber\\
	&=& {} - p\zeta \underbrace{\sum_{y=0}^{\infty} \Poisson_{\lambda+\zeta}(y)}_{=1} + p \underbrace{\sum_{y=0}^{\infty} \Poisson_{\lambda+\zeta}(y) y}_{=(\zeta + \lambda)} \log \left( 1 + \frac{\zeta}{\lambda} \right) \nonumber \\
	& & {} - p + p^2 - p^2e^{\frac{\zeta^2}{\lambda}} \nonumber \\
	&=& p(\zeta + \lambda) \log \left( 1 + \frac{\zeta}{\lambda} \right) - p\zeta - p - p^2\left( e^{\frac{\zeta^2}{\lambda}} - 1 \right).\IEEEeqnarraynumspace \label{eq:dark_l1}
\end{IEEEeqnarray}
For small enough $\Ave$, we choose $\zeta = \sqrt{\lambda\log\frac{1}{\Ave}}$ and $p = \frac{\Ave}{\zeta} = \frac{\Ave}{\sqrt{\lambda\log\frac{1}{\Ave}}}$. By using~\eqref{eq:lower_general} and~\eqref{eq:dark_l1} and letting $\Ave$ tend to zero we 
establish the lower bound in~\eqref{eq:dark}.


\section{the upper bounds}\label{sec:upper}

In this section we shall sketch the proofs of the upper bounds on the asymptotic capacities of the Poisson channel. We shall use the duality bound~\cite{lapidothmoser03_3} which states that, for any distribution~$R(\cdot)$ on the output, the channel capacity satisfies
\begin{equation}\label{eq:duality}
	C \le \sup \E{ D\bigl(W(\cdot|X) \| R(\cdot) \bigr)},
\end{equation}
where  the supremum is taken over all allowed input distributions. We shall describe the choices of $R(\cdot)$ that lead to our upper bounds, but we shall omit the details.


\subsection{Dark Current Proportional to $\Ave$}\label{sub:zero_upper}

In this subsection we shall sketch the proof for~\eqref{eq:zero_upper}. To this end, as in \cite{lapidothmoser07_sub}, we introduce the \emph{Poisson channel with continuous output} whose input $x$ is the same as the original Poisson channel, and whose output is $\tilde{y}\in \Reals_0^+$. The conditional density $\tilde{W}(\cdot|\cdot)$ is
\begin{equation}\label{eq:continuous_Poisson}
	\tilde{W}(\tilde{y}|x) = \Poisson_{\lambda+x}(\lfloor\tilde{y}\rfloor ).
\end{equation}
We denote the capacity of~\eqref{eq:continuous_Poisson} under Constraints~\eqref{eq:ave} and~\eqref{eq:peak} by $\tilde{C}(\lambda,\Ave,\Peak)$. It is shown in~\cite{lapidothmoser03_4} that
\begin{equation*}
	C(\lambda,\Ave,\Peak) = \tilde{C}(\lambda,\Ave,\Peak).
\end{equation*}
Thus, to prove~\eqref{eq:zero_upper}, it suffices to prove
\begin{equation}\label{eq:zero_upper_con}
	\varlimsup_{\Ave\downarrow 0}\frac{\tilde{C}(0,\Ave,\Peak)}{\Ave\log\frac{1}{\Ave}} \le 1, \quad \Peak\in(0,\infty].
\end{equation}
To this end, we choose the distribution $\tilde{R}(\cdot)$ on $\tilde{Y}$ to be of density
\begin{equation*}
	f_{\tilde{R}}(\tilde{y}) = \begin{cases} (1-p),& 0\le \tilde{y}<1 \\ p\cdot\frac{\tilde{y}^{\nu-1}e^{-\frac{\tilde{y}}{\beta}}}{\beta^\nu \Gamma(\nu, \frac{1}{\beta})},&\tilde{y}\ge 1,\end{cases}
\end{equation*}
where $\beta>0$ is arbitrary, $\nu\in(0,1]$ and $p\in(0,1)$ will be specified later, and $\Gamma(\cdot,\cdot)$ denotes the Incomplete Gamma Function given by
\begin{equation*}
\Gamma(a,\xi) = \int_{\xi}^{\infty} t^{a-1} e^{-t} \d t, \qquad \forall \, a, \, \xi \ge 0.
\end{equation*}
Applying~\eqref{eq:duality} on $\tilde{C}(0,\Ave,\Peak)$ with the above choice of $f_{\tilde{R}}(\cdot)$ in the place of $R(\cdot)$ and with the choice $\nu=\frac{1}{2}$ yields that, for every $p\in(0,1)$ and $\beta>0$
\begin{IEEEeqnarray*}{rCl}
	\lefteqn{C(0,\Ave,\Peak)  \le  \Ave \log\frac{1}{p} + \log\frac{1}{1-p} + \frac{\Ave}{\beta}} \nonumber\\
	&&~~~~~~~~~~ {}+\Ave\max\left\{ 0, \left(\frac{1}{2}\log\beta + \log \frac{\Gamma(\frac{1}{2},\frac{1}{\beta})}{\sqrt{\pi}} + \frac{1}{2\beta}\right)\right\}.
\end{IEEEeqnarray*}
Choosing $p=\frac{\Ave}{1+\Ave}$ in the above inequality and letting $\Ave$ tend to zero yield~\eqref{eq:zero_upper}.


\subsection{Constant Nonzero Dark Current}\label{sub:dark_upper}

In this subsection we shall sketch the proof of the upper bound in~\eqref{eq:dark}. We choose the distribution $R(\cdot)$ on the output $Y$ to be
\begin{equation*}
	R(y) = \begin{cases} e^{-\lambda} \frac{\lambda^y}{y !}, & y \in \{ 0, 1, \ldots, N-1 \} \\
	\delta (1-p)p^{y-N}, & y \in \{N, N+1, \ldots\},\end{cases}
\end{equation*}
where $N\in\Integers^+$ and $p\in(0,1)$ are constants to be specified later, and $\delta$ is a normalizing factor given by
	$\delta \triangleq \sum_{y=N}^\infty e^{-\lambda} \frac{\lambda^y}{y!}$.
We next apply~\eqref{eq:duality} to upper-bound $C(\lambda,\Ave,\Peak)$. Calculation (with repeated applications of the Chernoff bound) yields
\begin{IEEEeqnarray}{rCl}
	\lefteqn{C(\lambda,\Ave,\Peak)} \nonumber\\
	& \le & \left(N\log N + \frac{1}{12N} + \frac{1}{2}\log(2\pi N) + \log\frac{1}{1-p}\right) \nonumber\\
	& & {}~~~~\cdot \left(\frac{\Ave }{N-\sqrt{N}-\lambda}+\exp\left( N + N\log \lambda - N \log N\right) \right)\nonumber\\
	&& \phantom{\frac{1}{p} }\nonumber\\
	&& \phantom{\frac{1}{p}}\nonumber\\
	& & {} + \left(1+\log\frac{1}{p} + \log \lambda \right) \cdot \bigg( \Ave  + \frac{\lambda \Ave }{N -\sqrt{N} -\lambda} \nonumber\\
	&& {}~~~~~~~~~~~~+ \lambda \cdot e^{N-1-\lambda + (N-1)\log\lambda - (N-1)\log(N-1)} \bigg) \nonumber\\
	&& {} + \Ave  \cdot \left(1+\frac{\lambda}{N-\lambda} \right) \cdot \max \left\{ 0, \log\frac{1}{\lambda}\right\} \nonumber\\
	&& {} + \Ave  \cdot \frac{N\log\frac{N}{\lambda}}{N-\lambda}.\label{eq:dark_upper417}
\end{IEEEeqnarray}
For small enough $\Ave $, we choose $N = \left\lfloor \log\frac{1}{\Ave } \right\rfloor$ and let $p \in (0,1)$ have any fixed value that does not depend on $\Ave$. Applying these choices to~\eqref{eq:dark_upper417}  and taking the limit $\Ave\downarrow 0$ yield the second inequality in~\eqref{eq:dark}.

\bibliographystyle{IEEEtran}           
\bibliography{/Volumes/Data/wang/Library/texmf/tex/bibtex/header_short,/Volumes/Data/wang/Library/texmf/tex/bibtex/bibliofile}


\end{document}